%% file: gated-MoS2QD-ee-paper_may08_2020.tex
\documentclass[aps,prb,twocolumn,superscriptaddress,longbibliography]{revtex4-1}
\usepackage{graphicx}
\usepackage{amsmath}
\usepackage{bbold}
\usepackage[colorlinks=true, citecolor=blue, urlcolor=blue, linkcolor=blue,bookmarks=false,hypertexnames=true]{hyperref} 

\usepackage[english]{babel}

\newcommand{\Ket}[1]{\left| #1 \right\rangle}
\newcommand{\Bra}[1]{\left\langle #1 \right|}

\begin{document}
\graphicspath{}

\title{Valley and spin polarized broken symmetry states of interacting electrons in gated MoS$_2$ quantum dots}

\author{Ludmi\l a Szulakowska}
\affiliation{Department of Physics, University of Ottawa, Ottawa, Ontario, Canada K1N 6N5}

\author{Moritz Cygorek}
\affiliation{Department of Physics, University of Ottawa, Ottawa, Ontario, Canada K1N 6N5}

\author{Maciej Bieniek}
\affiliation{Department of Physics, University of Ottawa, Ottawa, Ontario, Canada K1N 6N5}
\affiliation{Department of Theoretical Physics, Wroc\l aw University of Science and Technology, Wybrze\.ze Wyspia\'nskiego 27, 50-370 Wroc\l aw, Poland}

\author{Pawe\l \ Hawrylak}
\affiliation{Department of Physics, University of Ottawa, Ottawa, Ontario, Canada K1N 6N5}

\date{\today}

\begin{abstract}
Understanding strongly interacting electrons enables the design of materials, nanostructures and devices. Developing this understanding relies on the ability to tune and control electron-electron interactions by, e.g., confining electrons 
to atomically thin layers of 2D crystals with reduced screening.  The interplay of strong interactions on a hexagonal lattice with  two nonequivalent valleys, topological moments, and the Ising-like spin-orbit  interaction gives rise to a variety of phases of matter corresponding to valley and spin polarized  broken symmetry states.
In this work we describe a highly tunable strongly interacting system of electrons laterally confined to monolayer transition metal dichalcogenide MoS$_2$ by metalic gates. We predict the existence of valley and spin polarized broken symmetry states tunable by the  parabolic confining potential using exact diagonalization techniques for up to $N=6$ electrons.
We find that the ground state is formed by one of two phases, either  both spin and valley polarized or valley unpolarised but
spin intervalley antiferromagnetic, which compete as a function of electronic shell spacing. This finding can be traced back to the combined effect of Ising-like spin-orbit coupling and weak intervalley exchange interaction. These results provide an explanation for interaction-driven symmetry-breaking  effects in valley systems and highlight the important role of  electron-electron interactions for designing valleytronic devices.
\end{abstract}


\maketitle

\section{Introduction}
The role of electron-electron (e-e) interactions in determining the many-electron ground and excited states in different materials is controlled by the ratio of Coulomb energy $V$ to kinetic energy $T$ as  $ V/T= r_s$ where $\pi {r_s}^2$ is an area per electron \cite{vignale_quantum_2005}. For small $r_s$  electrons in 2D are well described by the Fermi liquid theory but as $r_s$ increases, density decreases, and the spin polarised and Wigner crystal phases follow \cite{attaccalite_correlation_2002, zarenia_wigner_2017}. In the 2D Hubbard model the electronic phases are controlled by the ratio of on-site Coulomb energy to the tunneling matrix element $U/t$. On a hexagonal lattice,  calculations predict a semimetallic phase followed by the anti-ferromagnetic and Mott-insulating phase \cite{sorella_semi-metal-insulator_1992, otsuka_universal_2016, wehling_strength_2011}. The $U/t$ can be tuned by controlling screening ($U$) or controlling $t$. Recent work on twisted bilayer graphene (BG) showed that $t$ can be significantly reduced by twisting layers in BG \cite{luican_single-layer_2011}. The quenching of tunneling results in strongly correlated system with Mott-insulating and superconducting phases \cite{cao_correlated_2018, cao_unconventional_2018}. Recent experiments in BG \cite{nomura_quantum_2006, weitz_broken-symmetry_2010} 
and transition metal dichalcogenides (TMDCs)  point to potential existence of spin \cite{roch_spin-polarized_2019} and valley polarized \cite{scrace_magnetoluminescence_2015} interaction driven broken symmetry valley and spin polarised states. 

In this work we focus on a new emerging highly tunable strongly interacting system of $N$ electrons laterally confined to monolayer 2D crystal, such as MoS$_2$,  \cite{geim_van_2013, splendiani_emerging_2010, mak_atomically_2010, kadantsev_electronic_2012, yu_many-body_2019, van_tuan_coulomb_2018} by metalic gates. \cite{volk_electronic_2011, liu_intervalley_2014, kormanyos_spin-orbit_2014, guclu_graphene_2014, song_gate_2015, pawlowski_valley_2018, pisoni_gate-tunable_2018, bhandari_imaging_2018, wang_electrical_2018, chen_magnetic_2018, pawlowski_spin-valley_2019, kurzmann_excited_2019, lin_many-body_2019, bieniek_effect_2020, knothe_quartet_2020}
The confinement to a single atomic layer leads to reduced screening and enhanced e-e interactions manifested in large, $\sim300$ meV, exciton binding energies. \cite{mak_atomically_2010, qiu_nonanalyticity_2015, jadczak_robust_2017,  lin_many-body_2019, bieniek_band_2020}  Metallic gates can be used to define quantum dots (QDs) with 
discrete levels with spacings $\omega$ and enable a controlled charging of the QDs with $N$ electrons. The ratio $V/T$ scales with $\omega$ as  $ V/T= 1/\sqrt{\omega}$ . In small GaAs QDs at large $\omega$, the ground state (GS) is well approximated by configurations minimizing single particle (SP) energy \cite{korkusinski_designing_2003}, but in large QDs, for small $\omega$, spin polarised  \cite{hawrylak_single-electron_1993, mikhailov_quantum-dot_2002, korkusinski_designing_2003} and correlated phases emerge \cite{korkusinski_designing_2003, korkusinski_pairing_2004}.  Here we combine the atomistic multimillion atom description of SP states, lateral confining potential and realistic e-e interaction matrix elements with accurate exact diagonalization techniques to determine GS and excited states of electrons in MoS$_2$ QDs. 

\section{Model }

\begin{figure}[h]
\centering
\includegraphics[scale=0.28]{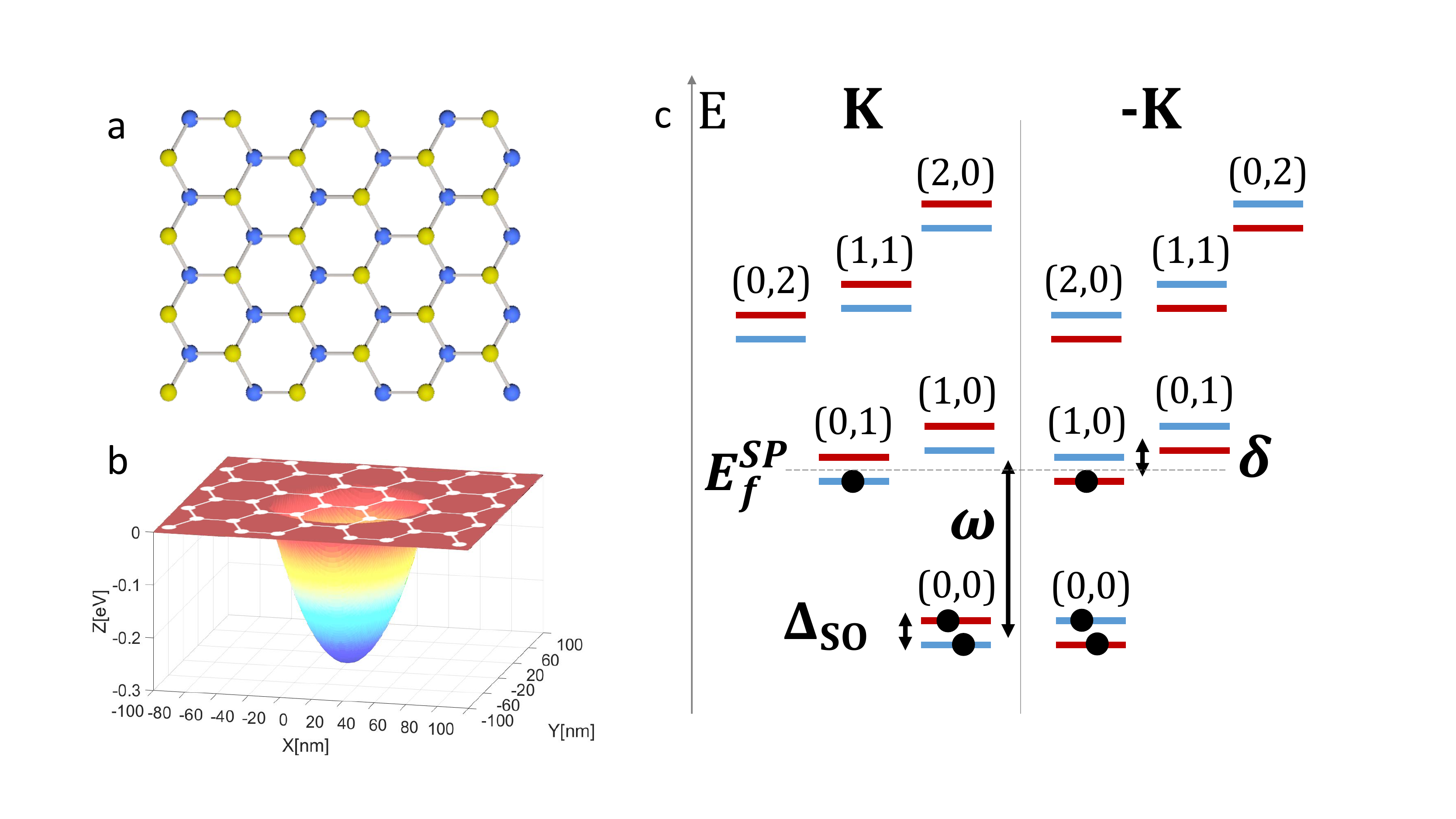}
\caption{\textbf{MoS$_2$ quantum dots}. a. Rectangular computation box of monolayer MoS$_2$. Blue (yellow) atoms denote Mo (S) atoms. b. An example of the parabolic confining potential from the metallic gates with a depth of $300 meV$. White lattice represents the computation box in which electrons are confined to form a quantum dot (QD). c. Single-particle energy structure of an MoS$_2$ QD. The harmonic oscillator (HO) shells are doubly degnerate due to valleys $K$ and $-K$. $\omega$ denotes shell spacing. Indices $(n,m)$ label HO states. Spin down and up energy levels are split by the spin-orbit splitting $\Delta_{SO}$, opposite in opposite valleys. HO shells are split by $\delta$ due to valley topological moments. To obtain a half-filling for 2 HO shells $N=6$ electrons are needed. $E_F^{SP}$ denotes the Fermi level for a non-interacting system of $N=6$ electrons. \label{fig1}}
\end{figure}

Fig. \ref{fig1}a shows the top view of a monolayer MoS$_2$ lattice. The blue (yellow) atoms correspond to Mo metal (S) atoms with 3 $d$-orbitals (3 $p$-orbitals) as described in detail in 
Ref. \cite{bieniek_effect_2020}. The single particle (SP) Hamiltonian describes the tunneling of electrons between Mo d-orbitals and S p-orbitals. The conduction band (CB) wavefunctions are expanded in Mo and S orbitals and computed for a large computational box with $\sim10^6$ atoms and periodic boundary conditions. The SP states are hence characterised by a band index and a wavevector $\boldsymbol{k}$, free from edge states present in a finite computational box.  The relevant  band structure consists of valence band (VB) and conduction band (CB), with the smallest energy gap at the two non-equivalent CB minima at $+K$ and $-K$ points of the Brillouin zone, and of six additional minima (3 per $K$-point), at the Q-points. 

In such a realistic computational box we add a gate-defined parabolic confining potential $V(r)$
(described in \ref{methods}), as shown in Fig. \ref{fig1}b. The confining potential mixes the CB states, lowers their energy into the energy gap of MoS$_2$ and confines electrons to the center of a QD. 

A schematic picture of a typical low energy SP spectrum is presented in Fig.~\ref{fig1}c. It is doubly degenerate due to the valley index $+K,-K$. In each valley the spectrum resembles that of a 2D harmonic oscillator (HO), consisting of shells of states separated by the spacing $\omega$, which can be tuned by the depth of the confining potential or  
the radius of the QD. \cite{hawrylak_single-electron_1993, raymond_excitonic_2004}
A separate six-fold degenerate HO-like spectrum 
originating from Q points is present at higher energies (not shown). These states are not occupied by electrons 
for the range of $\omega$ considered here and are not considered in what follows. 
The SP levels are labelled with quantum numbers $p, \sigma$, where $p=[(n,m),K]$ contains state index $(n,m)$ and valley index $+K,-K$, and $\sigma$ denotes spin. The $(n,m)$ are HO quantum numbers, 
where $n+m$ determines the shell index and $L=n-m$ is the angular momentum of the
state in valley $+K$ (left) or $-K$ (right). 
The spin-orbit induced Zeeman splitting $\Delta_{SO}$ between spins 
($\uparrow$ and $\downarrow$ shown in red and blue respectively
in Fig. \ref{fig1} c) is opposite for both valleys.

A further modification of a simple HO level structure is a topological splitting $\delta$ proportional to $\omega$, exhibited by all electronic shells. This splitting arises from the valley topological moments in each  valley. This results in 
opposite angular momentum $L$ states as the lowest energy states in $+K$ and $-K$ valleys.

We next turn to filling SP spectrum with electrons up to the Fermi level $E_f^{SP}$, as illustrated in Fig. \ref{fig1} c) for $N=6$ electrons. Our goal  is to understand the many-body GS and excited states of interacting electrons and explain how the interactions mix many configurations in forming correlated electronic states. 

The many-body Hamiltonian in the basis of SP QD states reads
\begin{equation}
H=\sum_{p\sigma}{e_{p\sigma} c_{p\sigma}^{\dagger} c_{p\sigma}\ }+\frac{\eta}{2}\sum_{pqst\sigma\sigma'}{\Bra{pq}V\Ket{st} c_{p\sigma}^{\dagger} c_{q\sigma'}^{\dagger} c_{s\sigma'}c_{t\sigma}}, \label{eq1}
\end{equation}
where the first term describes  energies $e_{p\sigma}$ of the SP HO states ${p\sigma}$ shown in \ref{fig1} c) and the second term describes interaction energy,  with Coulomb matrix elements describing scattering between pairs of states.
The Coulomb matrix elements ${\Bra{pq}V\Ket{st}}$ are computed using atomistic million atom orbitals and with both bare Coulomb and Keldysh-screened interaction \cite{rytova_screened_1967, keldysh_coulomb_1979}, accounting for reduced screening by 2D materials [see \ref{methods} for details]. The parameter $\eta$ allows us to turn the e-e interaction on and off.

The $N$-electron configurations are constructed as $\Ket{x}= \prod_{p,\sigma} {c_{p\sigma}}^{\dagger}\Ket{0}$,
 and the wavefunction of $N$-electron system is expanded in all possible electronic configurations $\Ket{x}$. The Hamiltonian matrix in the space of configurations $\Ket{x}$ is constructed and diagonalized giving exact eigenstates and eigenvalues. For example, for $N=6$ we find up to $\sim5\cdot10^7$ configurations for  $M=60$ SP orbitals.

 We now turn to discuss the properties of $N$-electron systems. We focus here on $N=2$ , $N=4$ and $N=6$ electrons. 
This is because in a non-interacting system filling the first s-shell  requires $N=4$ electrons. Half filling of the s-shell is realised with $N=2$ electrons and to obtain half filling of the first 2 shells  $N=6$ electrons are needed.

\section{Results}
\subsection{N=2 electron complex}
In order to build the understanding of the role of interactions in MoS$_2$ QDs, it is instructive to first focus on $N=2$ electrons on the first 4-fold degenerate s- shell of SP states. 
 In the absence of SO splitting $\Delta_{SO}$, this system describes the half-filled lowest energy shell of  BG QD \cite{volk_electronic_2011, kurzmann_excited_2019, knothe_quartet_2020}
or a half-filled  p-shell of a self-assembled QD \cite{korkusinski_designing_2003}. As we will see the GS is determined by the exchange interaction and can be understood in terms of spin singlets and  triplets.

\begin{figure}[h]
\centering
\includegraphics[scale=0.3]{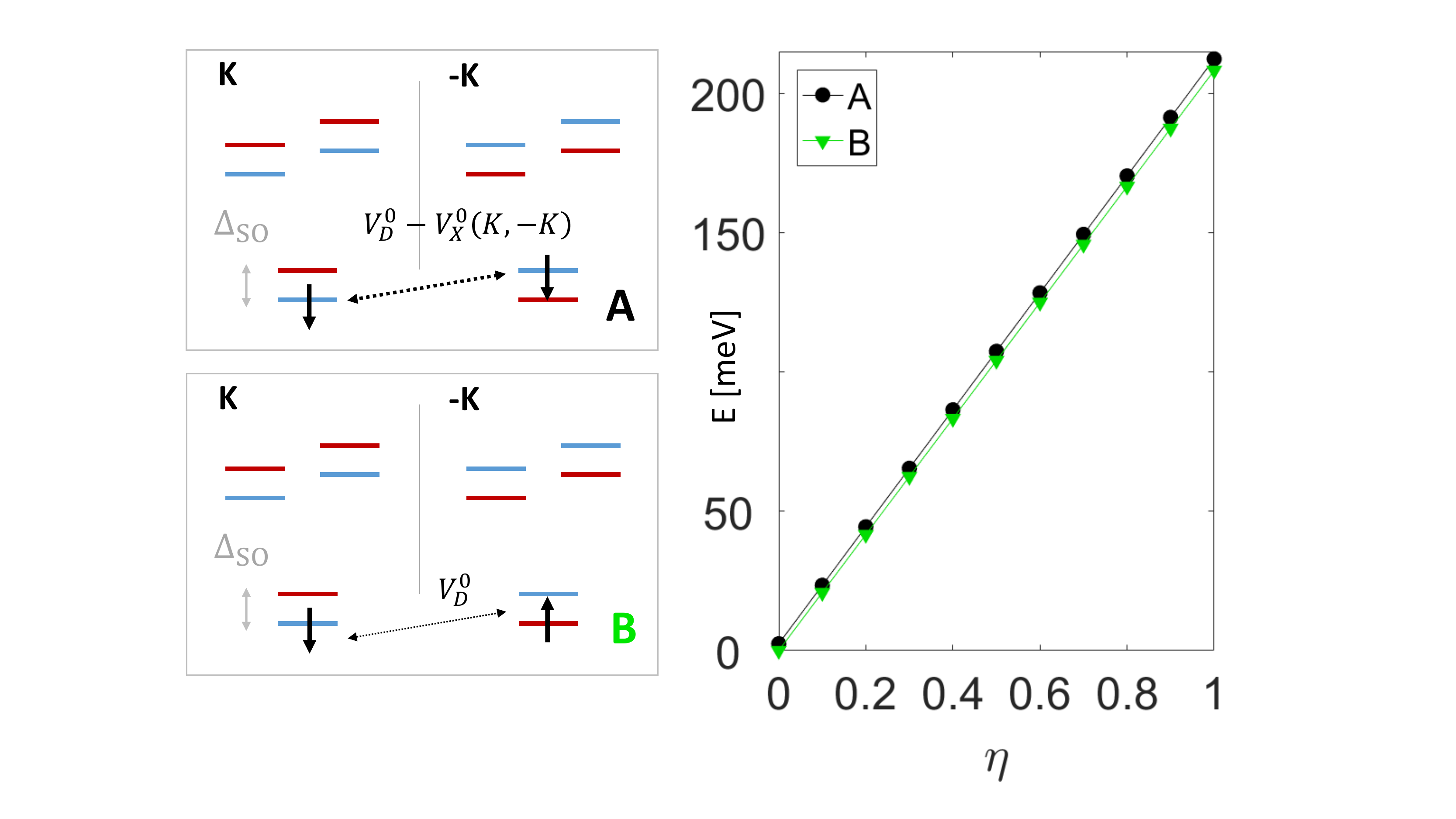}
\caption{$N=2$ electron configurations in the s-shell. Configurations A (spin polarised) and B (spin unpolarised) of $N=2$ electrons in the s-shell are valley unpolarised. Two electrons in A interact with direct interaction $V_D^0$ and intervalley exchange $V_X^0(K,-K)$ (both include Keldysh screening), but have a higher single particle energy due to SO splitting $\Delta_{SO}$. The electrons in B interact only with $V_D^0$. The right panel shows the energies of A and B for increasing strength of interaction $\eta$. Due to much smaller intervalley exchange compared to SO splitting $V_X^0(K,-K)\ll\Delta_{SO}$, B has always lower energy. B makes up the many-body GS for $N=2$ electrons in 1 shell, which is valley and spin unpolarised.  \label{fig2}}
\end{figure}

With $N=2$ electrons on s-shell orbitals  in opposite valleys the  $N=2$ electron spin  states can be classified into three spin triplets $\Ket{T^s_+}=\Ket{\uparrow}\Ket{\uparrow}$,  $\Ket{T^s_0}=\frac 1{\sqrt{2}}\big( \Ket{\uparrow}\Ket{\downarrow}+\Ket{\downarrow}\Ket{\uparrow}\big)$, or $\Ket{T^s_-}=\Ket{\downarrow}\Ket{\downarrow}$. The total wavefunction  is therefore simultaneously  a valley singlet $\Ket{S^v}=\frac 1{\sqrt{2}}\big(\Ket{K}\Ket{-K} - \Ket{-K}\Ket{K}\big)$. The spin triplet valley singlet state
 $\Ket{S^v}\Ket{T^s_-}$ is shown schematically in  Fig. \ref{fig2}A . In the absence of $\Delta_{SO}$, the energy of the spin triplet configuration $E_T$ is composed of the sum of SP energies of s-type $(0,0)$  orbitals, the direct interaction $V_D^0$ and intervalley exchange  $V_X^0(+K,-K)$  to give
  $E_T= e_{00,\uparrow}+ e_{00,\downarrow}+ V_D^0(+K,-K)-V_X^0(+K,-K)$. 
  The exchange interaction lowers the energy of $\Ket{T^s}$ compared to $\Ket{S^s}$.
Exact diagonalisation of the $N=2$ electron system on the lowest s-shell with $\Delta_{SO}=0$ gives $\Ket{S^v}\Ket{T^s}$ as the triply degenerate GS, due to the intervalley exchange $V_X^0(+K,-K)$. This is in accordance with what has been found for half filled  p-shell of QDs \cite{wojs_charging_1996} and for BG QDs \cite{kurzmann_excited_2019}, a two valley system with negligible SO.

In TMDCs the Ising-like SO interaction  leads to spin splitting in the CB ranging from $\sim3 meV$ in Mo-based material to $\sim30meV$ in W-based material \cite{kadantsev_electronic_2012, scrace_magnetoluminescence_2015}.  Turning on $\Delta_{SO}$ leads to a decrease in the  energy of spin-down states in valley $K$ as well as of spin-up states in valley $-K$ . For the case of $N=2$ electrons  this means that the spin triplets $\Ket{T^s}$ and spin singlets $\Ket{S^s}$ mix and the three-fold degeneracy of the $\Ket{S^v}\Ket{T^s}$ GS is broken by the $\Delta_{SO}$. 
The splitting $\Delta_{SO}$ competes with intervalley exchange  $V_X^0(+K,-K)$. For weak intervalley exchange $V_X^0(+K,-K)\ll\Delta_{SO}$ the spin unpolarized state, depicted as configuration B in Fig. \ref{fig2}, becomes the lower energy state separated by a gap from the spin polarized states, configuration A in Fig. \ref{fig2}. This is shown in Fig. \ref{fig2} (right) for $\omega=36meV$ and varied strength of Keldysh-screened Coulomb interactions  $\eta$. 
Configuration B can be written as a mixture of  $\Ket{S^v}\Ket{T^s_0}$ and $\Ket{T^v_0}\Ket{S^s}$ and becomes the spin-valley singlet $\Ket{S^{sv}}=\frac 1{\sqrt{2}}\big( \Ket{S^v}\Ket{T^s_0}-\Ket{T^v_0}\Ket{S^s}  \big)=\frac 1{\sqrt{2}}\big(\Ket{K\downarrow}\Ket{-K\uparrow}-\Ket{-K\uparrow}\Ket{K\downarrow} \big)$.

We now lower the level spacing $\omega$ and allow the second shell of p-type single-particle states to be occupied by a second electron, e.g. as shown in Fig. \ref{fig3}D and Fig. 1C in Supplementary Material. 
The transfer from the s-shell to the p-shell costs SP energy $\omega+\delta/2$ but it is compensated by gain in interaction energy. Instead of $V_D^{0}(+K,-K)-V_X^0(+K,-K)$ for configuration A, the interaction is now $ V_D^1(+K,+K)-V_X^1(+K,+K)$ (the superscript denotes $L$ of the second electron). This change lowers the energy of D compared to A and B. This is because of the significantly stronger intra-valley exchange $V_X^1(+K,+K)$ compared to inter-valley $V_X^0(+K,-K)$. There are two possible p-shell orbitals and two possible $N=2$ electron configurations, out of which D (with the second electron in  $L=+1$ ($L=-1$) orbital at $+K$ ($-K$)) is lower in energy, as discussed in Supplementary Material.

\begin{figure}[h]
\centering
\includegraphics[scale=0.3]{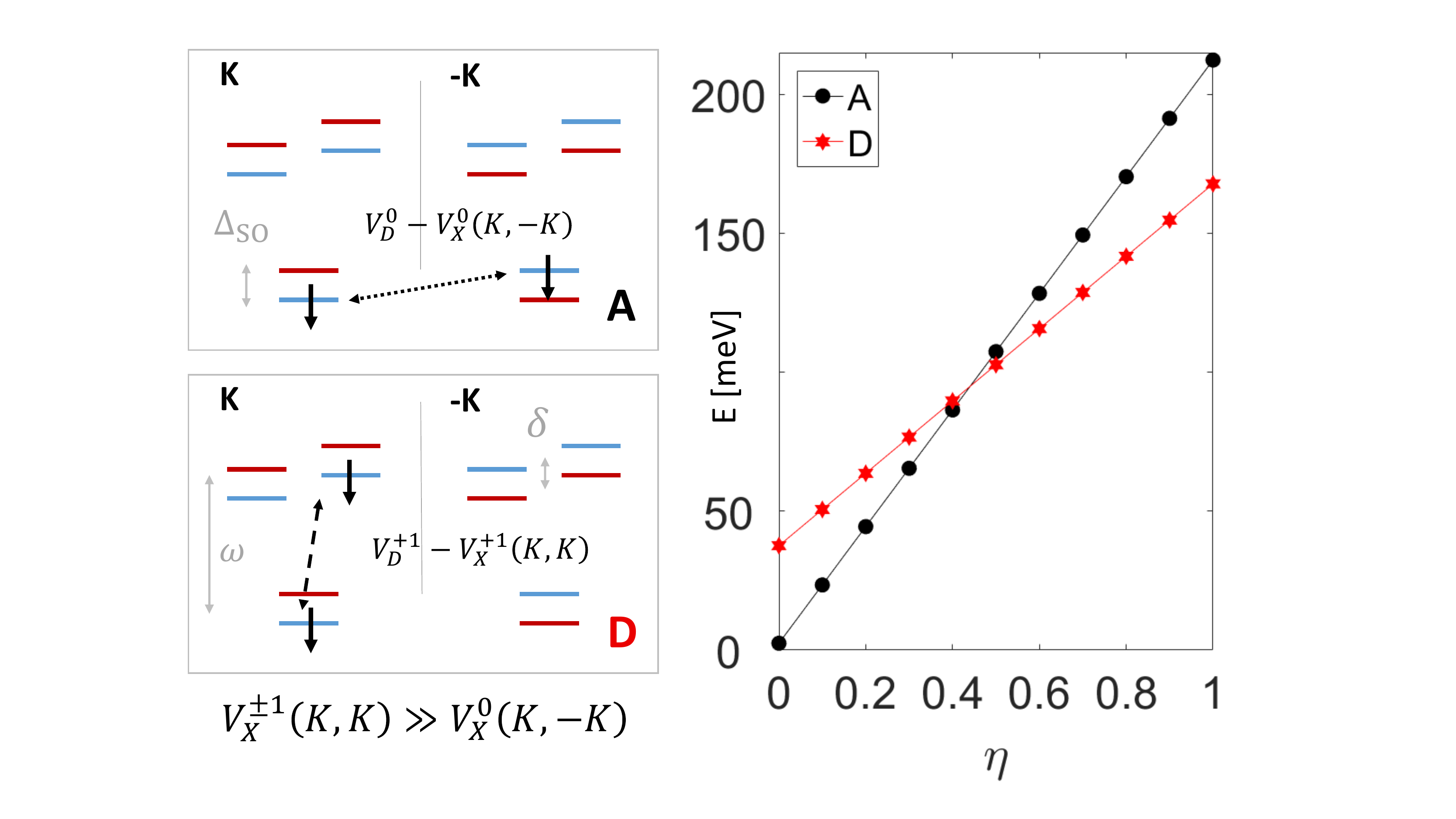}
\caption{$N=2$ electron configurations in 2 shells. Configurations A and D of $N=2$ electrons are both spin polarised, with no valley polarisation in A and full valley polarisation in D. The electrons in A interact with intervalley exchange $V_X^0(+K,-K)$, while the electrons in D interact with much stronger intravalley exchange $V_X^{+1}(K,K)$, which compensates the single-particle energy cost $\omega$. These competing energies are responsible for generating an energy transition visible in the right panel as a function of the strength of interaction $\eta$. For a non-interacting case, at $\eta=0$ A is lower in energy, but for an interacting system, D has lower energy. \label{fig3}}
\end{figure}

This competition between configurations A and D is shown for $\omega=36meV$ in  Fig. \ref{fig3} (right). For a non-interacting system, at $\eta=0$, the valley and spin polarised configuration D with one electron on a p-shell has higher energy compared to valley unpolarised configuration A. However, as strength of interactions $\eta$ is increased for Keldysh screened Coulomb interactions, we see a transition and the valley-spin polarised configuration D moves to lower energy.
This transition can also be understood by considering the $N=2$ electron wavefunction. Configuration D is a product of spin triplet $\Ket{T^s_-}$, valley triplet $\Ket{T^v_-}$ and hence electronic orbital s-p singlet $\Ket{S^e}$, written as $\Ket{D}=\Ket{S^e}\Ket{T^{sv}_-}$, where $\Ket{T^{sv}_-}=\Ket{T^v_-}\Ket{T^s_-}$ (corresponding $\Ket{T^{sv}_+}$ is degenerate). All other spin, valley and electronic orbital configurations can be constructed in a similar way, taking into account the nonzero $\Delta_{SO}$ (see Suplementary Materials for details). For higher shells D-like configurations $\Ket{T^{sv}_-}$ (valley-spin polarised) compete for the GS with the triplet configurations $\Ket{T^{sv}_0}$ (valley-spin unpolarised).

\subsection{GS and excited states of $N\geq2$ and $M=60$}

We have so far identified different possible phases of the $N=2$ electron system and different interactions competing to produce the GS and excited states: SP energies, SO splitting, topological moments, direct and exchange intravalley and intervalley interactions. We now describe results of exact diagonalisation of the $N=2-6$ electron problem as a function of $\omega$ for varying number of electronic shells. Converged results for 5 shells per valley ($M=60$ SP states)  for $N=2$ and $N=6$ electrons are discussed below and remaining electron numbers are discussed in Supplementary Material.    
All our numerical results show spin valley locking in the many-body GS,  with spin $\uparrow$ ($\downarrow$) electrons occupying valley $+K$ ($-K$)  so that $N_{\downarrow}=N_K$ and $N_{\uparrow}=N_{-K}$, which is in line with our explanation of the GS of the $N=2$-electron system.
This allows us to label the GS with one polarisation quantum number
$ \widetilde{V}=\frac{N_K-N_{-K}}{N}$,
denoting total valley polarisation and equal here to the total spin polarisation $\widetilde{V}=S_z\frac{2}{N}=\frac{N_{\uparrow}-N_{\downarrow}}{N}$. 

\begin{figure}[h]
\centering
\includegraphics[scale=0.28]{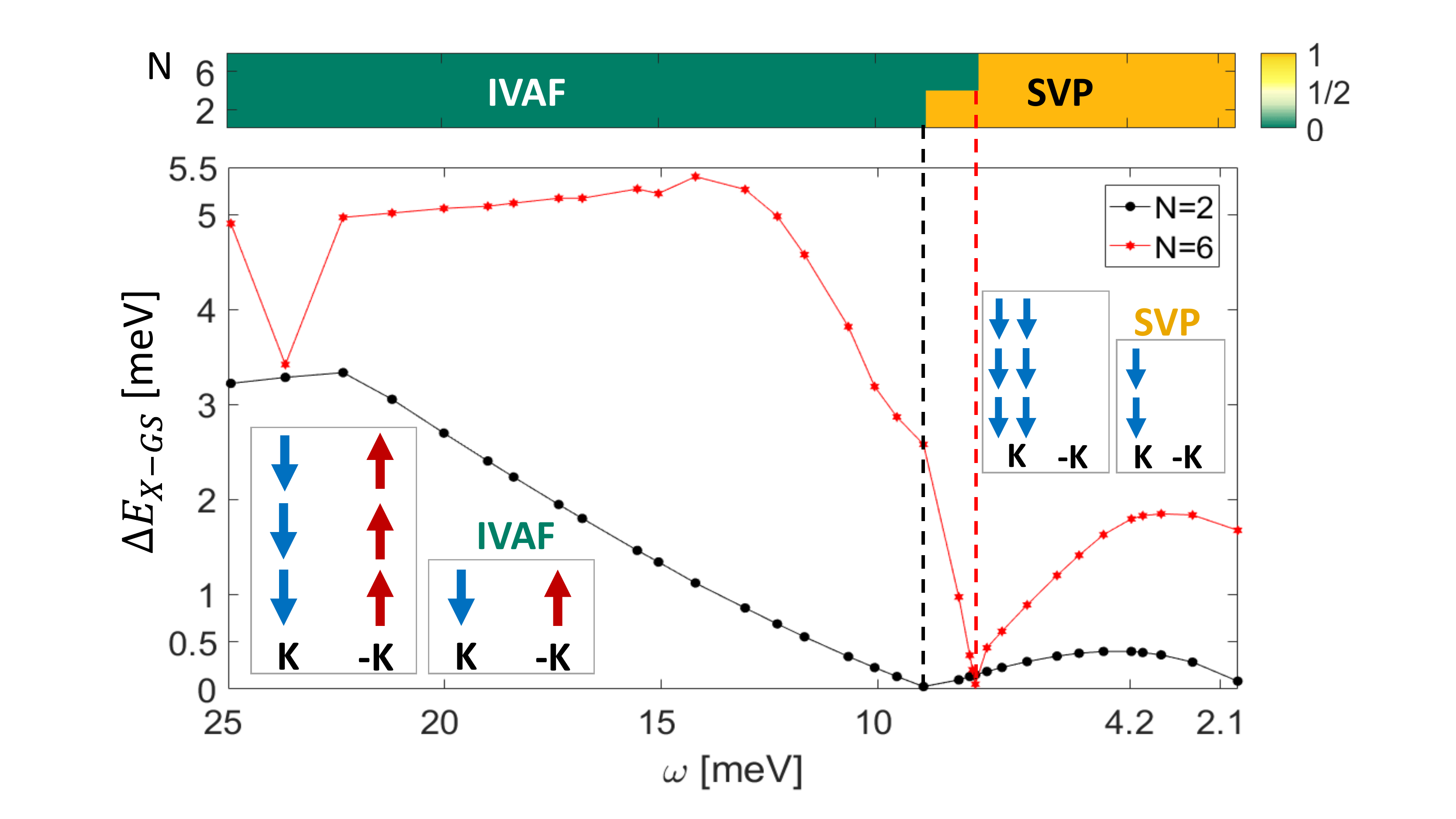}
\caption{Transition in the many-body ground state character with varying level spacing $\omega$. The top panel shows the nature of the ground state (GS) for $N=2$ and $N=6$ electrons as a function of the HO shell spacing $\omega$. Colors depict the polarisation quantum number $\widetilde{V}$ (see text for details): dark green depicts the intervalley antiferromagnetic (IVAF) GS while orange stands for spin and valley polarised (SVP) GS. The bottom panel contains the energy difference $\Delta E_{X-GS}$ between the GS and the first excited state for varying $\omega$ for $N=2$ (black dots) and $N=6$ (red stars). Vanishing $\Delta E_{X-GS}$ marks a transition in the nature of the GS from the IVAF to the SVP phase (shown with dashed lines). Insets give schematic representaiton of the IVAF and the SVP GS phases (red and blue arrows depict spin up and down).    \label{fig4}}
\end{figure}

The results for valley and spin polarisation $\widetilde{V}$ for $N=2$ and $N=6$ electrons are shown in the top panel of Fig.~\ref{fig4} while the corresponding energy gaps,  $\Delta E_{X-GS}=E_X-E_{GS}$ where $E_X$ is the first excited state,  and schematic electron configurations are shown in the lower panel.  

The colors in Fig.~\ref{fig4} (top) denote the degree of polarisation  $\widetilde{V}$: orange depicts full spin and valley polarisation (SVP) with total $|S_z|=N/2$,  while dark green identifies a fully inter-valley anti-ferromagnetic (IVAF) 
GS with total $S_z=0$ ($N_{\uparrow}=N_{\downarrow}$)  and no net valley polarisation ($N_K=N_{-K}$). Schematic configurations corresponding to IVAF and SVP phases are shown for both $N$.  Clear phase transitions from the  IVAF GS to the SVP GS  accompanying the closure of energy gaps $\Delta E_{X-GS}$ at critical energy spacings $\omega_C\approx9$meV and  $\omega_C\approx$8meV are visible for $N=2$ and $N=6$ electrons respectively. For $N=2$, the phases IVAF and SVP correspond to the competing triplets $\Ket{T^{sv}_0}$ and $\Ket{T^{sv}_{\pm}}$ respectively.

In the insets of Fig. \ref{fig3}. we show a schematic representation of the two competing GS phases for $N=2$ and $N=6$ electrons with spin $\uparrow$ ($\downarrow$) electrons shown with red up (blue down) arrows. IVAF (left) involves $N_{\uparrow}=N_{-K}=N_{\downarrow}=N_K=1$ and $N_{\uparrow}=N_{-K}=N_{\downarrow}=N_K=3$ for $N=2$ and $N=6$ respectively. The SVP phase (right) is fully polarised with $N=N_{\downarrow}=N_K=2$ (and a degenerate time-reversed state with $N=N_{\uparrow}=N_{-K}=2$) and similarly $N=N_{\downarrow}=N_K=6$ (and a degenerate time-reversed state with $N=N_{\uparrow}=N_{-K}=6$).

In order to detect the competing GS phases in an experiment, one needs to consider the stability of these phases. It is partly determined by the energy spacing between the GS and excited state $\Delta E_{X-GS}$, which in turn impacts transport measurement. Closing of the energy gaps due to phase transitions would affect the temperature dependence and high-source-drain Coulomb diamonds in transport.The computed energy gaps $\Delta E_{X-GS}$  as a function of $\omega$ for $N=2$ (black) and $N=6$ (red)  reach several meV. The quantum phase transitions bewteen IVAF and SVP phases occur when $\Delta E_{X-GS}=0$.

\section{Conclusions}
Using atomistic theory combined with  exact many-body diagonalisation tools we predict the existence of  broken-symmetry Spin and Valley Polarized (SVP) and InterValley AntiFerromagnetic (IVAF) electronic states of interacting electrons 
electrostatically confined in a parabolic QD in a single layer of MoS$_2$. These results highlight the important role of electron-electron interactions for designing valleytronic devices.


\input{methods}

\section*{Acknowledgments}
 L.S., M.C., M.B., and P.H. thank M. Korkusinski, Y. Saleem, A. Altintas, A. Dusko, J. Manalo,
 A. Luican-Mayer, A. Badolato, I. Ozfidan, L. Gaudreau, S. Studenikin and A. Sachrajda for
discussions. L.S., M.C.,  M.B., and P.H. acknowledge support from NSERC Discovery and QC2DM Strategic
Project grants as well as uOttawa Research Chair in Quantum Theory of Materials, Nanostructures and Devices. M.B. acknowledges financial support from National Science Center (NCN), Poland, grant Maestro No. 2014/14/A/ST3/00654. M.C. acknowledges support from the Humboldt Foundation. Computing resources from Compute Canada  are gratefully acknowledged.


\bibliography{bib1}


\end{document}

%% file: methods.tex
\section{Methods \label{methods}}
The hexagonal MoS$_2$ layer consists of two triangular lattices, one of Mo atoms and a second of $S_2$ dimer.  We write our Hamiltonian in the basis of $d$ orbitals of Mo atoms and $p$ orbitals of S$_2$ dimers, which are even with respect to the metal plane, as \cite{bieniek_effect_2020}
{\begin{multline}
{\hat{H}} ^{TB}=\sum_{i}{E_ic_i^{\dagger} c_i}+\sum_{<i,j>}\left(T_{ij}c_i^{\dagger} c_j+h.c.\right)\\
+\sum_{\ll i,j\gg}\left(W_{ij}c_i^{\dagger} c_j+h.c.\right), \label{eqA1}
\end{multline}}
where $c_i^{\dagger}$ creates an electron on state $i$ , and $i$ carries atom unit cell  index, orbital index and sublattice index. $E_i$ are onsite energies and $T$ ($W$) are $6\times6$ nearest neighbour (NN) (next nearest neighbour (NNN)) hopping matrices between sites. Energies $E_i$ include the parabolic potential $V_i$ generated by the gates (as shown in Fig. \ref{fig1} b)) on a site corresponding to index $i$, with $V_i=V(\boldsymbol{r}_i)=\left|V_{max}\right|/(R_{QD}^2)\cdot r_i^2-V_{max}$, for $\left|\boldsymbol{r}_i\right|\leq R_{QD}$ and $0$ elsewhere. $V_{max}$ is the depth of the potential and $R_{QD}$ is the radius of the QD. 

To avoid edge states in the energy gap we apply periodic boundary conditions, i.e., we wrap the finite computational box on  a torus with periodic boundary conditions. The Hamiltonian in Eq. (\ref{eqA1}) can now be written in the basis of Bloch states as 
{\begin{multline}
{\hat{H}} ^{TB}_{\boldsymbol{k}-basis}=\sum_{\boldsymbol{k}}\sum_{\alpha}E_{\alpha}a_{\boldsymbol{k}\alpha}^{\dagger}a_{\boldsymbol{k}\alpha}\\
+\sum_{\boldsymbol{k}}\sum_{<\alpha,\beta>}\left(e^{i\boldsymbol{k}\boldsymbol{d}_{\alpha,\beta}}T_{\alpha,\beta}a_{\boldsymbol{k}\alpha}^{\dagger}a_{\boldsymbol{k}\beta}+h.c.\right)\\
+\sum_{\boldsymbol{k}}\sum_{\ll\alpha,\beta\gg}\left(e^{i\boldsymbol{k}\boldsymbol{d}_{\alpha,\beta}}W_{\alpha,\beta}a_{\boldsymbol{k}\alpha}^{\dagger}a_{\boldsymbol{k}\beta}+h.c.\right)\\
+\sum_{\boldsymbol{k},\boldsymbol{k}'}\sum_{\boldsymbol{R},\alpha}\left(e^{i(\boldsymbol{k}-\boldsymbol{k}')\boldsymbol{R}}V_{\boldsymbol{R},\alpha}a_{\boldsymbol{k}\alpha}^{\dagger}a_{\boldsymbol{k}'\alpha}+h.c.\right), \label{eqA1k}
\end{multline}}
where  $\boldsymbol{R}$ is position of a cell  and $\alpha$ carries orbital and sublattice index.  $\boldsymbol{d}_{\alpha,\beta}$ is the NN or NNN vector between NN or NNN orbitals $\alpha,\beta$, and only the confining potential $V_{\boldsymbol{R}\alpha}$ mixes the $\boldsymbol{k}$-states \cite{liu_intervalley_2014}.

We diagonalise Eq. (\ref{eqA1k}) to obtain valley specific quantum dot states  $p,\sigma$. 
In the second quantization the many-body Hamiltonian in the basis of SP QD states $p,\sigma$ reads:
\begin{equation}
H=\sum_{p\sigma}{e_{p \sigma}   c_{p \sigma}^{\dagger} c_{p\sigma}\ }+\frac{\eta}{2}\sum_{pqst\sigma\sigma'}{\Bra{pq}V\Ket{st} c_{p\sigma}^{\dagger} c_{q\sigma'}^{\dagger} c_{s\sigma'}c_{t\sigma}}, \label{eqA2}
\end{equation}
where in the first term $e_{p \sigma} $ are energies of the SP HO states $p$ and the second terms includes Coulomb scattering  between these states,with $\eta$ controlling  the strength of the interactions. We express the Coulomb matrix elements $\Bra{pq}V\Ket{st}$ in Eq. (\ref{eqA2}) in the basis of atomic orbitals as $\Bra{pq}V\Ket{st}=\sum_{ijkl}{{A_i^p}^\ast{A_j^q}^\ast A_k^sA_l^t\Bra{ij}V\Ket{kl},}  $ where $A$ are solutions to the SP Hamiltonian in Eq. (\ref{eqA1}) and Eq. (\ref{eqA1k}). We include only onsite short-range integrals $\Bra{ii}V\Ket{ii}$ and the long-range part is taken as a classical Coulomb term. The Coulomb integrals are calculated using Coulomb potential with Keldysh screening, using the 2D Fourier transform, as \cite{bieniek_band_2020}
{\begin{multline}
V_K^{3D}\left(\boldsymbol{r}-\boldsymbol{r}'\right)=\frac{1}{\epsilon^\ast}\frac{e^{2}}{4\pi\epsilon_0}\frac{1}{\left(2\pi\right)^2}\cdot\\
\cdot\int_{-\infty}^{\infty}\frac{2\pi}{\left|\boldsymbol{k}\right|}\frac{1}{1+2\pi\alpha\left|\boldsymbol{k}\right|}e^{-\left|z-z'\right|\left|\boldsymbol{k}\right|}e^{i\boldsymbol{k}\left(\boldsymbol{\rho}-\boldsymbol{\rho}'\right)}d^2\boldsymbol{k},
\end{multline}}
where $\alpha=2.2 \AA$ is the 2D polarisability and we take $\epsilon^*=2.5$.